\documentclass{emulateapj}

\usepackage{graphicx}
\usepackage{epstopdf}

\newcommand{\etc}{etc.}
\newcommand{\fig}[1]{Fig.~\ref{#1}}

\newcommand{\ie}{i.e.}

\newcommand{\flux}{erg cm$^{-2}$ s$^{-1}$  \AA$^{-1}$}

\newcommand{\htwo}{H$_2$}
\newcommand{\hetwo}{\ion{He}{2}}

\newcommand{\none}{\ion{N}{1}}
\newcommand{\ntwo}{\ion{N}{2}}

\newcommand{\oone}{\ion{O}{1}}
\newcommand{\otwo}{\ion{O}{2}}

\newcommand{\osix}{\ion{O}{6}}

\newcommand{\fuse}{{\it FUSE}}
\newcommand{\hst}{{\it HST}}

\newcommand{\iue}{{\it IUE}}

\slugcomment{To Appear in {\em Publications of the Astronomical Society of the Pacific}}

\shorttitle{HUT: The Final Archive}
\shortauthors{Dixon et al.}

\begin{document}


\title{The Hopkins Ultraviolet Telescope: The Final Archive}


\author{William V. Dixon}
\affil{Space Telescope Science Institute, 3700 San Martin Drive, Baltimore, MD 21218; dixon@stsci.edu}

\author{William P. Blair}
\affil{Department of Physics and Astronomy,
Johns Hopkins University, 3400 N. Charles Street, Baltimore, MD 21218; wpb@pha.jhu.edu}

\author{Jeffrey W. Kruk}
\affil{NASA/Goddard Space Flight Center, 8800 Greenbelt Road, Greenbelt, MD 20771; Jeffrey.W.Kruk@nasa.gov}

\and

\author{Mary L. Romelfanger}
\affil{Department of Physics and Astronomy,
Johns Hopkins University, 3400 N. Charles Street, Baltimore, MD 21218; mary@pha.jhu.edu}

\begin{abstract}
The Hopkins Ultraviolet Telescope (HUT) was a 0.9 m telescope and moderate-resolution ($\Delta \lambda = 3$ \AA) far-ultraviolet (820--1850 \AA) spectrograph that flew twice on the space shuttle, in 1990 December (Astro-1, STS-35) and 1995 March (Astro-2, STS-67). The resulting spectra were originally archived in a non-standard format that lacked important descriptive metadata.  To increase their utility, we have modified the original data-reduction software to produce a new and more user-friendly data product, a time-tagged photon list similar in format to the Intermediate Data Files (IDFs) produced by the {\it Far Ultraviolet Spectroscopic Explorer} calibration pipeline.  We have transferred all relevant pointing and instrument-status information from locally-archived science and engineering databases into new FITS header keywords for each data set.  Using this new pipeline, we have reprocessed the entire HUT archive from both missions, producing a new set of calibrated spectral products in a modern FITS format that is fully compliant with Virtual Observatory requirements.  For each exposure, we have generated quick-look plots of the fully-calibrated spectrum and associated pointing history information.  Finally, we have retrieved from our archives HUT TV guider images, which provide information on aperture positioning relative to guide stars, and converted them into FITS-format image files.  All of these new data products are available in the new HUT section of the Mikulski Archive for Space Telescopes (MAST), along with historical and reference documents from both missions.  In this paper, we document the improved data-processing steps applied to the data and show examples of the new data products.  
\end{abstract}


\section{Introduction}

The Hopkins Ultraviolet Telescope (HUT;  \citealt{HUT1CAL1}) was a 0.9 m telescope and moderate-resolution ($\Delta \lambda = 3$ \AA) far-ultraviolet (FUV) prime-focus Rowland-circle spectrograph developed and built by the Johns Hopkins University (JHU) under the leadership of Professor Arthur F. Davidsen.  HUT and two other instruments, the Ultraviolet Imaging Telescope (UIT; \citealt{Stecher:97}) and the Wisconsin Ultraviolet Photo-Polarimetry Experiment (WUPPE; \citealt{Nordsieck:94}), made up the Astro Observatory, a Spacelab payload that flew on two space-shuttle missions, STS-35 in December 1990 (the Astro-1 mission, on {\em Columbia}), and STS-67 in March 1995 (Astro-2, on {\em Endeavour}).  The three telescopes were co-aligned and shared a pointing system.  The Astro-1 mission also included the Broad-Band X-Ray Telescope (BBXRT; \citealt{Serlemitsos:92}), which was mounted on a pointing system separate from that of the UV telescopes.  The telescopes flew as an attached payload and were operated jointly by the astronauts and a large cadre of support scientists and engineers on the ground.

HUT was a ground-breaking instrument in several ways.  In first order, its 820--1850 \AA\ bandpass provided access to sub-Lyman $\alpha$ wavelengths (912--1216 \AA) unavailable to either the {\em International Ultraviolet Explorer (IUE)} or the {\em Hubble Space Telescope (HST)} and offered greater sensitivity and resolution than the short-wavelength band of \iue\ in the region of overlap. In second order, HUT observed a pair of local white-dwarf stars at extreme ultraviolet wavelengths (415--925 \AA; Astro-1 only). HUT was used to observe a broad range of targets, including solar system objects, cataclysmic variable stars, supernova remnants and planetary nebulae, nearby galaxies and starbursts, and quasars and active galaxies \citep{Davidsen:93}.  A NASA Astrophysics Data System search shows over 140 refereed publications using HUT data, placing the Astro Spacelab missions among the most scientifically productive space-shuttle missions ever flown.

The HUT spectra from both missions were archived initially at the National Space Science Data Center (NSSDC) but were later moved to what is now the Mikulski Archive for Space Telescopes (MAST).  This version of the archived data was lacking in several important respects.  While stored in a valid FITS format, they used an IRAF-inspired structure that required special software to interpret.  Important instrument parameters, including the aperture position angle and any pointing offsets from the principal target position, were not included among the file-header keywords, leading to possible confusion for archival users.  Finally, while HUT raw-data files provide time resolution of either 1 ms or 2 s, depending on the count rate, no photon arrival-time information was preserved in the original data products, which were stored in a cumulative data format.

To address these shortcomings, we have re-processed the entire HUT archive from both Astro missions, modifying the original data-reduction software to produce a new data product, a time-tagged photon list. Similar in format to the Intermediate Data Files (IDFs) produced by the {\em Far Ultraviolet Spectroscopic Explorer (FUSE)}\/ calibration pipeline \citep{Dixon:07}, these files record the arrival time and detector coordinate of each photon event recorded by the HUT detector.  We have transferred all relevant pointing and instrument-status information from archived science and engineering databases at JHU into new FITS header keywords.  We have added routines to correct bit-flip errors in the IDFs and to flag data obtained when the target was out of the spectrograph aperture.  We have written software to extract and fully calibrate spectra directly from the IDFs and used it to produce a new set of calibrated spectral products in a modern FITS format.  All of our data products are fully compliant with Virtual Observatory requirements \citep{Tody:2012}.  These newly processed data have been archived and are available through the \anchor{http://archive.stsci.edu/hut}{HUT pages at MAST}.

We provide three additional data products. The first is a quick-look overview plot for each exposure, showing the fully-calibrated spectrum and the pointing error and target count rate throughout the exposure.  These plots are available through the MAST browser interface.  The second are HUT TV guide-camera images of the target field, obtained at regular intervals throughout each exposure, which are useful for confirming ancillary information such as the location of guide stars and the aperture orientation.  The third are the fully-calibrated first-order spectra of the white dwarfs G191-B2B and HZ~43 obtained on Astro-1, from which second- and third-order flux has been subtracted.  These spectra are provided as high-level science products.  Finally, all relevant historical and technical documentation about the HUT project has been transferred to MAST from the retired JHU HUT project web site, providing a lasting on-line repository of reference information for archival users.

In Section \ref{sec_instrument} of this paper, we review the properties of the HUT instrument and its observing modes.  In Section \ref{sec_idf}, we step through the HUT calibration pipeline and describe how the results of each calibration step are recorded in the IDF.  Section \ref{sec_manipulate} describes our new software to correct errors in the IDF and flag times when the target was out of the aperture.  Section \ref{sec_extract} explains how the calibrated spectral file is constructed from the IDF.  The new HUT video frames are described in Section \ref{sec_video}, the quick-look plots in Section \ref{sec_quick}, and the high-level science products in Section \ref{sec_high}.  The contents of the new HUT web site, hosted by MAST, are described in Section \ref{sec_web}.  Section \ref{sec_summary} presents a summary of our efforts.  The formats of the IDF, extracted spectral files, and HUT TV files are detailed in Appendix \ref{sec_formats}.  A handful of peculiar targets are described in Appendix \ref{sec_notes}.

\begin{deluxetable}{cll}
\tablecolumns{3}
\tablewidth{0pt}
\tablecaption{HUT Apertures\label{apertures}}
\tablehead{
\colhead{Number} & \colhead{Astro-1} & \colhead{Astro-2}}
\startdata
1	&	30\arcsec\ diameter	&	12\arcsec\ diameter \\
2	&	9\farcs4 $\times$ 116\arcsec	&	32\arcsec\ diameter \\
3	&	30\arcsec\ diameter, Al filter	&	32\arcsec\ diameter, Al filter \\
5	&	17\arcsec\ $\times$ 116\arcsec, CaF$_2$ filter	&	19\arcsec\ $\times$ 197\arcsec \\
6	&	17\arcsec\ $\times$ 116\arcsec	&	10\arcsec\ $\times$ 56\arcsec \\
7	&	18\arcsec\ diameter	&	20\arcsec\ diameter
\enddata
\end{deluxetable}

\section{The HUT Instrument}\label{sec_instrument}

HUT consisted of a 0.9 m mirror and a prime-focus Rowland-circle spectrograph. First-order spectral coverage ranged from 820 to 1850 \AA.  The resolution ranged from 2 to 4.5 \AA, with minima near 900 and 1600 \AA, where the flat plane of the detector crossed the curved focal surface of the spectrograph (see Figs.\ 20--22 of \citealt{HUT2CAL2}).  The detector was a microchannel-plate intensified one-dimensional Reticon diode array.  Because of this design, no intrinsic spatial information (perpendicular to the dispersion axis) was available.  The plate scale at the detector was 2.8\arcsec\ \AA $^{-1}$, so when the pointing was stable ($\pm 1$\arcsec), the resolution was not significantly degraded by image motion.  To allow observations of bright targets, the telescope doors could
be partially closed, exposing 3.9\%, 14.6\%, or 50\% of the (5120
cm$^2$) primary mirror.  For the brightest targets, the doors could
be fully closed and smaller doors with areas of 1 cm$^2$ or 50
cm$^2$ could be opened.  Both long-slit and circular apertures were available, although the sizes of these apertures were different on the two Astro missions (Table \ref{apertures}).

The configuration and performance of HUT on the Astro-1 and Astro-2  missions are described in \citet{HUT1CAL1} and \citet{HUT2CAL1}, respectively.  \citet{HUT1CAL2, HUT2CAL2} present the final calibration of HUT for each mission.  For Astro-1, the HUT optics were coated with iridium, which provided about 20\% reflectivity in the both the extreme (500--600 \AA) and far (912--1216 \AA)  ultraviolet.  The Astro-1 mission was beset by a number of technical and hardware glitches, including the pointing system performance.  In all, 86 targets were observed by HUT, although many were observed multiple times.  Using an aluminum filter that blocked first-order light, HUT also made unique and important observations of two hot white dwarfs in second order, G191-B2B and HZ~43.  For the Astro-2 mission, the HUT optics were coated with silicon carbide (SiC), which provided a reflectivity of 30\% to 40\% over the full first-order bandpass.  With a longer mission and better technical performance of the entire system, HUT was used to observe 343 unique targets.  Again, many were observed multiple times, either to increase the signal-to-noise or to monitor time-variable objects.  No second-order observations were obtained on Astro-2, as SiC has a poor reflectance at EUV wavelengths.  

HUT, UIT, and WUPPE were co-aligned and mounted on the Spacelab Instrument Pointing System (IPS), which operated as an attached payload.  Observations were accomplished by a multi-stage process that involved the proper orientation of the shuttle, the positioning of the telescope pointing system, and the fine acquisition and guiding of the telescopes themselves. The astronauts pointed the IPS using video from the HUT TV guide camera (\S\ \ref{sec_video}).  Controlling the IPS with a joystick, the crew adjusted its position to move the target into the aperture, using fiducials that showed the expected positions of guide stars for each observation.  Once the target was acquired, the IPS used guide-star centroids computed by HUT to maintain pointing.  As mentioned above, technical problems on Astro-1 often prevented the auto-guide feature from working, and pointing was accomplished by manual control of the IPS throughout each exposure by the astronauts.  On Astro-2, the system worked flawlessly, and the resulting data are of significantly higher quality.

For scientific observations, HUT was operated in either of two observing modes, both of which produced a raw data file every two seconds.  In high-time-resolution mode, the file contained a photon-event list, reporting the detector X (dispersion direction) coordinate and arrival time (with an accuracy of $\pm$1 ms) of each photon.  This mode was limited to sources whose total count rates did not exceed $\sim$ 600 counts s$^{-1}$.  For brighter sources, a histogram mode was employed, producing a cumulative 2048-pixel histogram every two seconds.  By subtracting one raw histogram from the next, the number of photons arriving in each pixel during a two-second period could be determined.

\section{Intermediate Data Files (IDFs)} \label{sec_idf}

In order to provide HUT data in a more user-friendly format, we have adopted the time-tagged photon list format of the Intermediate Data Files (IDFs) employed by CalFUSE, the data-reduction pipeline for {\em FUSE}\/ \citep{Dixon:07}.  We have modified the format, however, for two reasons:  First, while the \fuse\/ data are two-dimensional, the HUT data are not; only a photon's X coordinate was recorded.  Second, we assumed that \fuse\/ users would have access to the entire CalFUSE data-reduction package, including its various calibration files.  While it is possible to retrieve a version of the HUT calibration pipeline from MAST, we do not expect users of these new data products to do so.  Instead, we record all of the necessary calibration information within each IDF.  The  format of the HUT IDF is detailed in Appendix \ref{idf_format}.

We began with the HUT data-reduction pipeline described by \citet{HUT2CAL2}.  The pipeline consists of a series of programs, written in C, that collect the raw data files, group them into individual exposures, construct raw-counts spectra, correct for various instrumental effects, and finally produce fully flux- and wavelength-calibrated spectra.  Each routine operates as before, but now calls a new subroutine that creates or modifies the appropriate IDF.  We have transferred all relevant pointing and instrument-status information from locally-archived science and engineering databases into new FITS header keywords.  We have added new routines to identify and separate offset pointings into separate exposures, correct for telemetry bit flips in the data stream, and flag times when a target was outside the spectrograph aperture. 

\subsection{Generating the IDF}

The initial pipeline module reads a set of raw data files, in either histogram or time-tag format, and produces spectra with units of raw counts per pixel.  This task is complicated by the fact that the Astro Observatory was controlled in real time by astronauts aboard the space shuttle.  They had the freedom to change instrument parameters, such as the door state or spectrograph aperture, or to tilt the telescope mirror to observe a secondary target during an observation.  Support scientists on the ground could also program such changes into the observing sequences uploaded to the telescope, which would execute them automatically.  As a result, a single planned observation may yield spectra of multiple offset positions observed with different instrument parameters. The pipeline tracks telemetry for the position of the telescope mirror, the aperture wheel, the telescope doors, \etc, producing a separate raw spectrum for each change of instrument parameters.  Each time it writes a raw spectrum to disk, it calls a new subroutine, {\tt cf\_ttag\_init}, which constructs a corresponding IDF.

In preparation for flight, a database file was constructed for each planned observation, detailing target parameters, instrument configuration, and expected guide stars.  We used these files to populate header keywords within each IDF, including the target's coordinates, magnitude, and redshift, the aperture position angle, and numerous comments describing the goals of the observation.  It was not uncommon to use motions of the primary mirror to move a target out of the aperture and obtain a spectrum of the nearby background.  To identify such offset pointings, we search for multiple exposures based on the same database file but with mirror offsets that differ by more than 6\arcsec.  For such offset exposures, the string ``{\tt \_offset}'' is appended to the target name, and the headers include the actual coordinates of the pointing position, rather than those of the primary target.
 
As described in Appendix \ref{idf_format}, the IDF contains two arrays of status flags, each of which employs the bit codes listed in Table \ref{flags}.  The first set of status flags is associated with the Photon Event List, which is stored in the first extension of the IDF.  At this stage of the pipeline, only the airglow bit is set, and only for photons whose X coordinates place them near one of these bright airglow features:
\otwo\ $\lambda 834$,
\oone\ $\lambda 989$,
Lyman $\beta$ $\lambda 1026$,
\none\ $\lambda 1134$,
\hetwo\ $\lambda 584$ (second order),
Lyman $\alpha$ $\lambda 1216$,
\oone\ $\lambda 1304$,
\oone\ $\lambda 1356$,
\otwo\ $\lambda 834$ (second order), and
\hetwo\ $\lambda 584$ (third order).

The second array of status flags is associated with the Timeline Table (second extension), which contains one entry for each raw data file.  In this array, the day/night flag is set to 0 if the data were obtained during orbit night, 1 if during orbital day.  The limb-angle flag is set whenever the angle between the target and the earth limb falls below 15\arcdeg, as absorption by \ntwo\ in the atmosphere can depress spectra shortward of 950 \AA\ at limb angles up to 12\arcdeg\ \citep{HUT2CAL2}.  Because the Astro-1 mission took place during solar maximum, charged-particle fluxes during passage through the South Atlantic Anomaly (SAA) were unacceptably high.  The Astro-2 mission occurred during solar minimum, and particle fluxes during the SAA were considerably lower.  Observations of bright targets, for which the higher background would contribute negligibly to the total signal, were scheduled during SAA passes on Astro-2.  Thus, Astro-1 data obtained during SAA passes are flagged as bad; Astro-2 data are not.  The high-background flag is set whenever the total count rate between 850 and 895 \AA\ exceeds 10 counts s$^{-1}$.  The target-out-of-aperture flag is set later in the pipeline and will be discussed below.

The various calibration steps are controlled by a set of file-header keywords initially set to either ``PERFORM'' or ``OMIT.''  By default, all keywords are set to ``PERFORM.''  If the exposure is identified as an airglow observation, then the AP-FLAG, JT-FLAG, and PH-FLAG keywords are set to ``OMIT,'' turning off the target-out-of-aperture test, the correction for telescope motion, and the photometric correction.  One could argue that these steps should be omitted for all diffuse targets (such as supernova remnants), but we have done so only for the observations of Comet Levy and the planetary nebula NGC 1535; these objects are discussed in Appendix \ref{sec_notes}.

\begin{deluxetable}{ll}
\tablecolumns{2}
\tablewidth{0pc}
\tablecaption{Bit Codes for IDF Status Flags\label{flags}}
\tablehead{
\colhead{Bit} & \colhead{Value}}
\startdata
        8 & User-defined bad-time interval \\
        7 & Not used \\
        6 & Target out of aperture \\
        5 & Background unacceptably high \\
        4 & Observation during SAA (Astro-1 only) \\
        3 & Limb-angle violation \\
        2 & Airglow feature \\
        1 & Day/Night flag (N = 0, D = 1) \\
\enddata
\tablecomments{Flags are listed in order from most- to least-significant bit.}
\end{deluxetable}

\subsection{Correcting for Telescope Motion}\label{sec_imcs}

Small telescope motions during an exposure have two effects on the data: first, as the target drifts in the dispersion direction, its spectrum shifts on the detector; the resulting smearing slightly reduces the spectral resolution.  We refer to this effect as jitter.  Second, as the target drifts near or beyond the edge of the aperture, flux is lost.  A pair of routines were developed to correct for these effects; both are described in Appendix A of \citet{HUT1CAL2}.

The first program, called {\tt imcs\_ph\_corr}, combines pointing information from the observatory (the Image Motion Compensation System, or IMCS) with the observed target count rate (excluding airglow features) to calculate the pitch and yaw errors throughout the exposure, the (integral) shift in pixels necessary to correct the observed spectrum for these drifts, and the effective throughput of the exposure.  The program now calls the new subroutine {\tt cf\_imcs\_corr}, which writes photometric information (a quality flag, the best-fit mean target count rate, and the number of data frames used in the analysis, among others) to keywords in the primary file header.  The pixel shifts are applied to the XRAW array and the results written to the X array in the Photon Event List.  Pointing information (a time series of pixel shifts, pitch errors, and yaw errors) is written to the IMCS Table (fourth extension) as described in Appendix \ref{idf_format}.

{\it Caveats:}  {\tt imcs\_ph\_corr} attempts to fit a model combining pointing information, the instrument point-spread function (assumed to be a two-dimensional Gaussian with a FWHM of 4\farcs5), and the dimensions of the aperture to the observed target count rate.  If the reduced $\chi^2$ value of the best-fit model is less than 10 (an arbitrary but effective value), then it is considered a success, and its corrections for jitter and photometric losses are applied to the data.  If pointing information is unavailable for 100 s or more during an exposure, then the jitter correction is not applied.

The second program, {\tt poisson\_corr}, compares the observed count-rate distribution of the target  with a Poisson distribution of the same mean count rate using the Kolmogorov-Smirnov test.  The program iteratively discards periods with low count rates until the probability that the distribution is Poisson is acceptably high.  A quality flag and the corrected mean count rate are written to the IDF file header via the new subroutine {\tt cf\_poisson}.

\subsection{Detector Dead Time}

If photons strike the detector faster than they can be read out by the detector electronics, counts may be lost.  Corrections for these dead-time effects are calculated by performing a detailed Monte Carlo simulation of the detector response for each input spectrum.  This procedure is described in Appendix B of \citet{HUT1CAL2}.  The program produces a pair of 2048-element arrays containing the dead-time correction and its associated error for each spectral element.  These arrays are passed to the new subroutine {\tt cf\_deadtime}, which writes them to the second extension of the IDF, called the Calibration Table.

\subsection{Phosphor Persistence}

As described by \citet{HUT2CAL2}, a phosphor-persistence effect causes approximately 8\% of photon events to be read twice by the detector electronics.  Because the effect is manifest as a constant multiplicative factor, it was absorbed into the definition of the effective area during the initial processing of the Astro-2 data.  For Astro-1, the raw-counts spectra were rescaled in a separate calibration step.  Here, we adopt the Astro-1 convention for reprocessing both data sets.  The phosphor-persistence correction, a scalar, is written as a header keyword to the IDF.

\subsection{Detector Dark Rate}

On Astro-1, the mean detector dark rate, measured on-orbit with the aperture closed, was $3.94 \times 10^{-4}$ counts s$^{-1}$ pixel$^{-1}$.  On Astro-2, the mean dark rate rose from $4.11 \times 10^{-4}$  to $4.76 \times 10^{-4}$ counts s$^{-1}$ pixel$^{-1}$ when the phosphor voltage setting was raised from 3 to 4 midway through the mission \citep{HUT2CAL2}.  The dark rate appropriate for each data set is stored in the IDF as a header keyword.  Orbit-to-orbit variations in the background were typically 20\%; these variations are addressed by the scattered-light correction (see below).

\subsection{Flat Field}

Flat-field features are fluctuations in the sensitivity of the instrument on wavelength scales of a few \AA ngstroms.  The flat field varied over the course of each flight, as airglow lines and bright emission-line targets reduced the gain in localized regions of the micro-channel plates.  For the Astro-1 mission, a single flat-field correction is available for each door configuration (both open, one closed, \etc).  For Astro-2, a set of time-dependent files is used to track the evolution of the flat field for each door configuration over the length of the mission.  The appropriate flat-field array is written to the Calibration Table of each IDF.

\subsection{Scattered Light and Second-order Correction}

The scattered-light and second-order corrections are performed not on the photon list, but on the extracted spectrum.  The scattered light is simply the average number of counts per pixel between 850 and 895 \AA\ for Astro-1, 840 and 885 \AA\ for Astro-2, a wavelength range free from airglow emission (and presumably from astrophysical emission as well, due to the neutral hydrogen cut-off at 912 \AA).  The scattered-light level is calculated on the fly and subtracted from each pixel in the extracted spectrum by the spectral-extraction routine.

To correct for the presence of second-order light at long wavelengths, we scale the spectrum between 905 and 932 \AA\ by the ratio of the second- to first-order efficiency (0.223 for Astro-1, 0.21 for Astro-2) and subtract it from the spectrum between 1810 and 1864 \AA.  The lower-wavelength limit was selected under the assumption that the interstellar medium is opaque below 912 \AA; this assumption may not be appropriate for  nearby white dwarfs and solar-system objects.  The upper limit represents the long-wavelength limit of the HUT bandpass.  Note that the spectral dispersion in second order is twice that in first order, so each second-order pixel maps to two first-order pixels.  The second-order correction is performed on the fly by the spectral-extraction routine.

\subsection{Wavelength Calibration}

For Astro-1 spectra, the HUT wavelength scale is assumed to be linear, beginning at 827.36 \AA\ and increasing by 0.51336 \AA\ per pixel.  For Astro-2, the wavelength scale is similar, but deviates from linearity by as much as 1.5 \AA\ \citep{HUT2CAL2}.  The appropriate wavelength-calibration array is written to the Calibration Table of the IDF, from which it is copied directly into the wavelength array of the extracted spectrum.

\subsection{Flux Calibration}\label{sec_fluxcal}

The HUT flux calibration is derived by comparing model atmospheres of hot DA white dwarfs with observations of white dwarf stars from each mission.  Details for the Astro-1 calibration are given by \citet{HUT1CAL2} and for Astro-2 by \cite{HUT2CAL2}.  The inverse-sensitivity curve represents the ratio of a synthetic stellar spectrum to the observed count-rate spectrum of the same star.  Its formal units are (\flux) per (count s$^{-1}$ pixel$^{-1}$), but they can be expressed more compactly as erg cm$^{-2}$ \AA$^{-1}$.  For Astro-1, a single set of inverse-sensitivity curves is available.  For Astro-2, a series of time-dependent curves spans the length of the mission; the curve for each exposure is derived by linear interpolation.  For both missions, separate curves are available for each configuration of the telescope doors.  The appropriate inverse-sensitivity curve is written to the Calibration Table of the IDF.

No inverse-sensitivity curve is available for the second-order spectra of G191-B2B (HUT109702) and HZ~43 (HUT105202) obtained on Astro-1.  The first-order curves written to their IDFs should not be used.  The spectra extracted from these IDFs have units of counts per second.

\subsection{Uncorrected Effects}\label{sec_uncorr}

The prime-focus design of the HUT spectrograph resulted in significant astigmatism in the cross-dispersion (detector Y) direction.  This effect was a function of wavelength, with the longest wavelengths most affected.  The size of the Reticon diodes was such that, for point sources, no flux was lost; however, for diffuse sources filling the long rectangular apertures, a small amount of flux could be missed.  Ray traces computed after the Astro-1 mission showed that, for apertures 116\arcsec\ long, this effect was $\sim$ 2\% at 1550 \AA, $\sim$ 4\% at 1650 \AA, and $\sim$ 6\% at 1850 \AA.  Correction factors for the three largest spectrograph apertures used on Astro-2 are presented in Fig.~17 of \citet{HUT2CAL2}.  Because this effect is small and its relevance depends upon the scientific goals of the observer, we have chosen not to correct for it in our reprocessing.

\section{Manipulating the IDFs}\label{sec_manipulate}

Once the time-dependent photon list is recorded in the IDF, we can manipulate the data in ways that are not possible with an extracted, one-dimensional spectrum.  We have written new routines to repair bit flips in the data stream, identify times when the target was out of the aperture, and copy the final set of status flags from the timeline table into the photon-event list.

\subsection{Repair Bit Flips}

HUT produced a raw data file every two seconds.  These files were recorded on the shuttle and transmitted to the ground at intervals throughout the mission.  Errors in the recording or transmission of data could result in bit flips, \ie, the transformation of a single bit from a zero to a one, or {\it vice versa.}  In histogram mode, such a flip is manifested as a bright (or faint) pixel in a single two-second frame.  Because our photon list is constructed by subtracting sequential histogram frames, such a bright pixel generates an entry with a large number of counts in a single pixel.  The next frame, without such a peak, generates an entry with a large, negative number of counts at the same pixel location.  To correct for this effect, we scan each photon list, identify negative-valued entries and the corresponding positive-valued peak, and add and subtract powers of two until the negative-valued trough is filled in.  This redistribution of counts may result in occasional photon-list entries with zero counts.  Because the total counts in each pixel are preserved, this exercise does not alter the final, extracted spectrum, but it greatly increases the utility of count rates derived from the data.  Accordingly, once any bit flips are repaired, the source, airglow, and background rate arrays in the timeline table are recomputed.
\clearpage
\subsection{Target Out of Aperture}

The source count rate (away from airglow lines) is stored in the timeline table (HDU 4) of the IDF.  We use these data to identify and flag times when the target was out of the aperture.  Excluding these times improves the signal-to-noise ratio of the extracted spectrum, principally by reducing the background contribution due to scattered airglow emission.  In general, the algorithm works as follows: the count-rate array is smoothed with a five-point boxcar.  If the peak value of the smoothed array is greater than 10 counts s$^{-1}$, then a threshold is set at 10\% of the peak.  If the peak is less than 10 but greater than 5 counts s$^{-1}$, then the threshold is set at 2 counts s$^{-1}$.  Elements of the smoothed array with values less than the threshold are flagged as bad.  If the peak is less than 5 counts s$^{-1}$, then no elements are flagged.  For bright targets, the process is repeated with the threshold raised to 20\% of the peak at the beginning and end of an exposure and for array elements adjacent to those already flagged as bad.

This algorithm works reasonably well, but many additional bad-time intervals were added by hand -- generally at the start of an exposure -- and a few intervals initially flagged as bad by the software were later deemed to be acceptable.  A few noteworthy cases are discussed in Appendix \ref{sec_notes}.  This correction is performed on diffuse targets, if they are bright enough, but not on exposures identified as airglow observations.

\subsection{Screen Photons}

Once the out-of-aperture flags in the timeline table are populated, the status flags in the timeline table are copied into the corresponding entries in the photon-event list.  The total and night-only exposure times are computed and written to the IDF header keywords EXPTIME and EXPNIGHT.

\section{Extracted Spectral Files}\label{sec_extract}

We have written a new program, {\tt cf\_extract}, to extract and calibrate spectra directly from the IDF.  The most straightforward approach to this task would be to construct a spectral histogram using the X coordinates (corrected for telescope motion) of each photon event, then apply each of the calibration steps discussed above.  Unfortunately,  the flat-field correction is defined in raw-pixel coordinates and cannot be applied to data already corrected for telescope motion.  This requirement leads to the following iterative scheme:  We step through the photon-event list in time order, accumulating a temporary spectral histogram (using the XRAW coordinate) and checking the difference between XRAW and X.  When this difference changes, we know that the telescope has moved, so we stop accumulating data.  The temporary spectral histogram is then corrected for detector dead time and phosphor persistence.  (The dead-time correction is smoothed by 13 pixels, so may be applied to either raw or corrected coordinates.)  The dark count is subtracted and the flat-field correction applied.  The spectrum is shifted from the XRAW to the X coordinate frame and added to a permanent spectral histogram.  This loop repeats until the photon list is exhausted.  

Because of dead-time and phosphor-persistence effects, the HUT electronics systematically miscounted the number of photon events recorded by the detector.  The shot noise in the observed spectrum, however, depends on the actual number of photon events.  We thus apply both the dead-time and persistence corrections, to obtain the true number of photons that hit the detector, before constructing a temporary variance array (equal at this point to the temporary spectral histogram).  The uncertainties associated with the persistence, dark-count, and flat-field corrections are assumed to be negligible and are ignored.  The temporary variance array is shifted and combined with the permanent variance array.  Once the permanent variance array is complete, we add to it the uncertainty associated with the dead-time correction.  (Again, the dead-time uncertainties are smoothed, so may be applied to either raw or corrected coordinates.)  The remaining calibration steps are performed on the permanent spectral histogram: subtraction of  scattered light and second-order counts, division by the exposure time, and multiplication by the appropriate inverse-sensitivity curve.  

If the photometric-correction keyword is set to ``PERFORM,'' then the pipeline must select between the IMCS and Poisson correction factors.  As described in \S\ \ref {sec_imcs}, the IMCS correction is considered a success if it returns a reduced $\chi^2$ value less than 10.  If so, then the IMCS correction is applied.  If not, then the Poisson correction is considered.  If its quality flag is unity, then the algorithm was successful and its correction applied.  If not, then no photometric correction is performed.  

This process is performed twice: the first time, all good data from the exposure are included.  The second time, only good data obtained during orbital night, as determined from the day/night flag in the status array, are used.  In low-earth orbit, the residual atmosphere produces strong emission features, particularly in the resonance lines of neutral hydrogen and oxygen.  This emission fills the aperture, and because the HUT apertures are relatively large, the resulting lines may span several \AA ngstroms on the detector.  These features are strongest during orbital day, when scattered Lyman $\alpha$ dominates the instrumental background.  On the night side of the orbit, the hydrogen lines are considerably weaker, and the oxygen lines essentially disappear.  For studies of emission-line sources or stellar absorption-line profiles, night-only spectra may be more useful than daytime or total spectra, despite the shorter exposure time.  An example from the \citet{Blair:95} analysis of the Puppis A supernova remnant is shown in \fig{fig_pupa}.  Both spectra are written to a single output file using the FITS format described in Appendix \ref{spec_format}.

For each HUT exposure, both IDF and extracted spectral files are available from MAST.  We do not provide software to modify the flags within an IDF, but creating such a tool would be straightforward.  Using {\tt cf\_extract}, the user may extract a fully-calibrated spectrum from a modified IDF, rejecting all photons flagged as bad.  The program is available for download from the MAST web site, along with instructions for its installation and use.  The program employs subroutines from the \anchor{http://heasarc.gsfc.nasa.gov/docs/software/fitsio/fitsio.html}{CFITSIO} subroutine library \citep{Pence:99}, which must also be installed.  

\begin{figure}
\epsscale{1.3}
\plotone{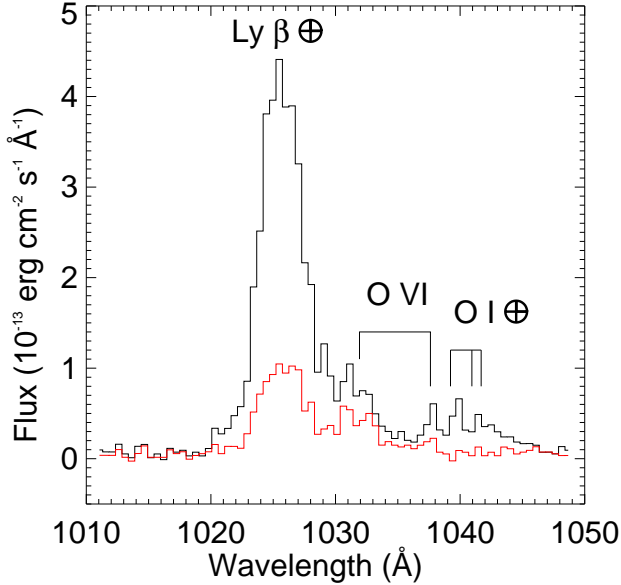}
\caption{HUT spectrum of a single filament within the Puppis A supernova remnant.  The black curve represents data from the entire 2960 s exposure.  The red curve represents data from the 2098 s obtained during orbital night.  The Lyman $\beta$ and \oone\ $\lambda 1040$ airglow features are considerably weaker in the night-only spectrum, revealing the \osix\ $\lambda 1032, 1038$ emission from the SNR.}
\label{fig_pupa}
\end{figure}

\section{HUT Video Frames}\label{sec_video}

The spectrograph apertures were etched into a mirrored surface that reflected visible light from the surrounding star field into a silicon intensified target (SIT) vidicon camera \citep{HUT1CAL1}.  The camera recorded a 9\farcm5 $\times$ 12\farcm3 field of view with 2\arcsec\ resolution.  Video images were digitized to 512 $\times$ 480 pixels and overlaid with fiducial marks that indicated the position and size of the selected spectrograph aperture and the positions of the guide stars pre-selected for each observation.  The instantaneous cumulative histogram of the spectrum was overplotted at the bottom of the display for reference.   

\begin{figure}
\epsscale{1.0}
\plotone{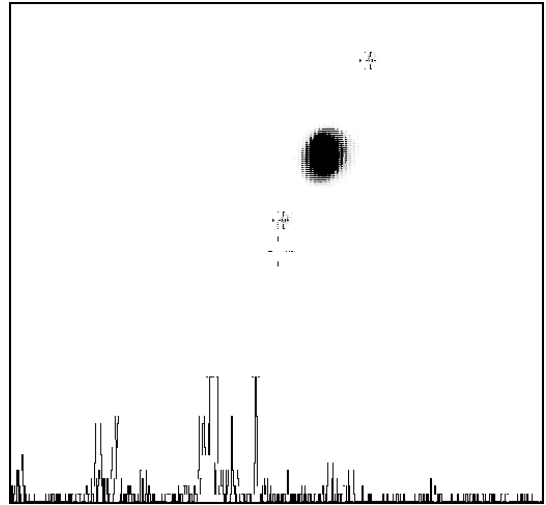}
\caption{HUT video frame from an observation of Jupiter's moon Io.  The field of view is 9\farcm5 $\times$ 12\farcm3.  Jupiter is clearly visible, and its moons Ganymede (lower left) and Europa are marked for use as guide stars.  Io itself is centered in the 20\arcsec\ aperture (indicated by a cross at the center of the image), so is not visible.  A cumulative histogram of Io's FUV spectrum (including the Io torus) is plotted across the bottom of the image.}
\label{fig_video}
\end{figure}

The HUT guider was the primary tool by which the astronauts verified the telescope pointing, performing the function of a TV guider at a ground-based telescope.  Selection of the appropriate camera gain, integration time, and neutral-density filter allowed acquisition of targets with visual magnitudes ranging from $-4$ to +17.  The rate at which TV images were saved (on the ground) varied with other demands on data storage, but was typically between one and four images per minute.  An example may be seen in \fig{fig_video}.  These images have never been archived, and users have not had access to them until now.

We have retrieved representative raw video frames for each exposure, converted them into FITS image files, and included complete astrometric information in their headers.  Note that there may be multiple TV frames associated with a single exposure.    These images will allow future users of HUT data to confirm visually the location and orientation of the spectrograph aperture relative to the target and guide stars at intervals throughout each exposure.

\section{Quick-Look Plots}\label{sec_quick}

\begin{figure*}
\centering
\plotone{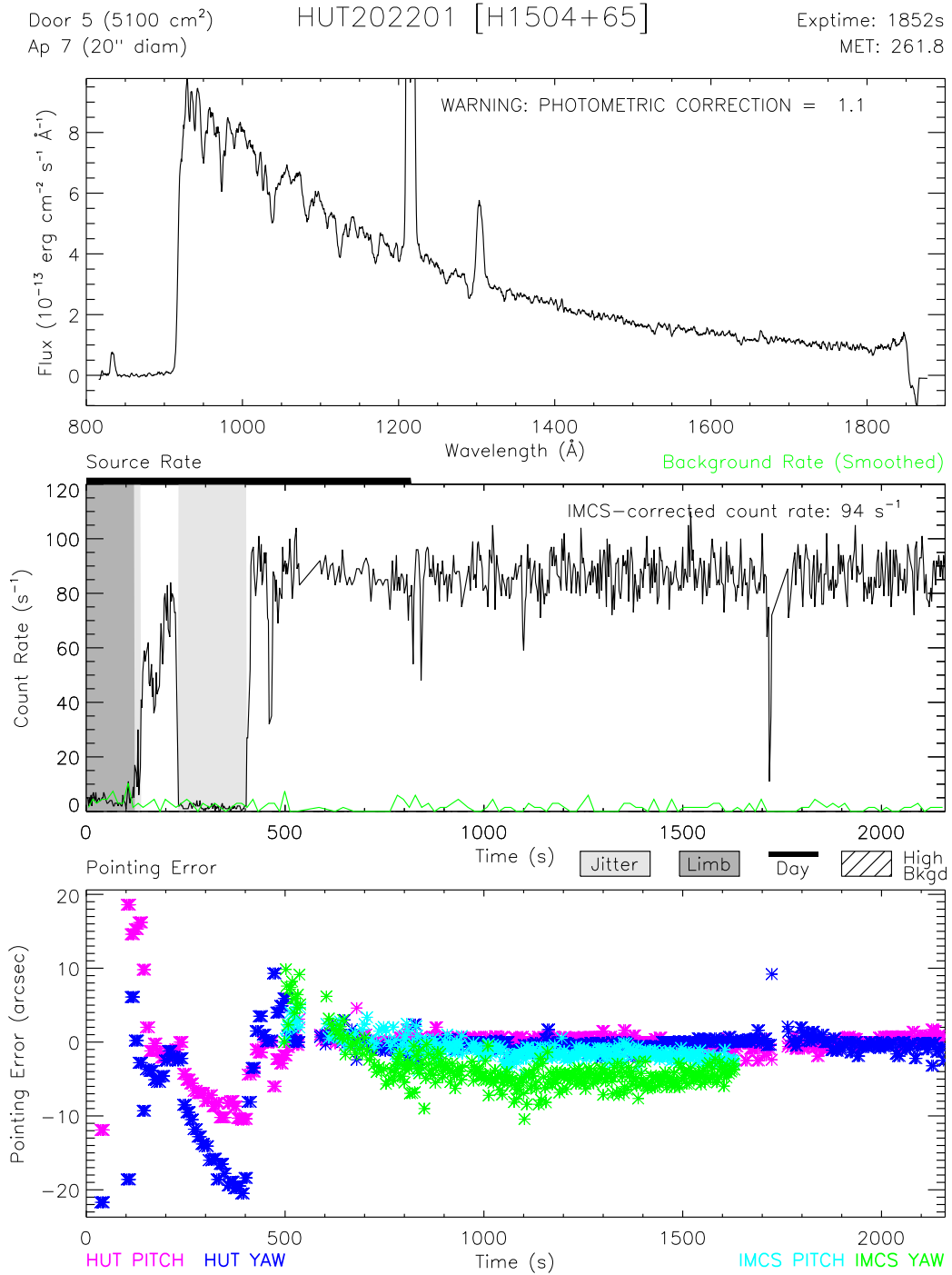}
\caption{Quick-look plot for exposure HUT202201, an observation of the white dwarf H1504+65.
{\em Top panel:}\/ Fully-calibrated spectrum of the star.
{\em Middle panel:}\/ Count-rate plots of the source (black) and background (green).  Data taken less than 15\arcdeg\ from the earth limb, when the star was out of the aperture, and during orbital day are marked with dark grey bands, light grey bands, and a black horizontal stripe, respectively.
{\em Bottom panel:}\/ Pitch and yaw pointing errors computed from the HUT guide stars (purple and blue) and by the Image Motion Compensation System (cyan and green).}
\label{fig_quick_look}
\end{figure*}

A quick-look plot is produced for each exposure.  \fig{fig_quick_look} presents the plot for exposure HUT202201, an observation of the white dwarf H1504+65.  The top panel shows the reprocessed, fully-calibrated spectrum of the star.  In the middle panel are plotted the count rate of the source (in black) and the background (in green).  Times when the target was less than 15\arcdeg\ from the earth limb are marked with a dark grey band, times when the star was out of the aperture are marked with a light-gray band, and times corresponding to orbital day are marked with a black horizontal stripe.  In the bottom panel, the pitch and yaw pointing errors computed from the HUT guide stars (purple and blue) and by the IMCS (cyan and green) are shown.

Let us investigate this figure in more detail.  The star was observed with both telescope doors open (door state 5) through the 20\arcsec\ spectrograph aperture.  The integration time, after bad times are discarded, is 1852 seconds.  The exposure began 261.8 hours after launch (MET = Mission Elapsed Time), while the star was less than 15\arcdeg\ above the earth limb.  Pointing errors in the first 500 seconds of the exposure moved the star out of the aperture, as can be seen in both the count-rate and pointing-error plots.  The remaining good-time intervals include enough time with the target near the edge of the aperture (around $t$ = 200 s) that the photometric correction, derived from the IMCS pointing errors, is 10\%, triggering a warning in the top panel.  Note the momentary drop in the count rate just after 1700 s, which corresponds to an excursion in the HUT YAW error in the bottom panel.

For each Astro-1 observation, G. Kriss provided a one-sentence description of the data quality for use on the HUT web site.  For Astro-1 data, we have included these capsule reviews in the IDF file headers under the keyword ASSESSMT.  If present, this comment is reproduced beneath the dataset name in the quick-look plots.

These quick-look plots are available for the reprocessed data from both missions as part of the MAST interface to the HUT archive.  Users can glean much information directly from these plots before deciding which data sets to download for use.

\clearpage
\section{High-Level Science Products}\label{sec_high}

Because the iridium-coated optics used on the Astro-1 mission were equally reflective at EUV and FUV wavelengths, the FUV spectra of the nearby white dwarfs G191-B2B and HZ~43 are contaminated by EUV flux in second and third order.  We have modeled and subtracted this flux.  The spectrum of G191-B2B is described in \S\ 2.2 and plotted in Fig.~2 of \citet{HUT1CAL2}.  The spectrum of HZ~43 is described in \S\ 5 and plotted in Fig.~8 of the same paper.  These spectra have been provided to MAST as high-level science products.

The optics used on Astro-2 were coated with SiC, which is much less reflective at EUV wavelengths.  The corrections for EUV flux are negligible for Astro-2 observations, so the standard pipeline products can be used without further correction.

\section{The HUT Archival Web Site at MAST}\label{sec_web}

Until recently, MAST provided supporting information about HUT and the HUT data products simply by pointing to the HUT web site at JHU.  Because former HUT team members have remained actively engaged at JHU, the original HUT web site has been maintained in an active but largely static mode since project funding ceased in 1996.  As part of the current project, we have transferred all relevant materials from the extant web site to a new archival web presence hosted by \anchor{http://archive.stsci.edu/hut}{MAST}.  We have scanned various technical documents, historical information, photographs, and other supporting materials and made them available through the new web site.

The MAST site provides access, through their standard interface, to the new intermediate data files, the extracted spectral files, the associated video frames, and the quick-look plots for each HUT exposure obtained on Astro-1 and Astro-2.  The extracted spectral files are made available to the Virtual Observatory through MAST's SSAP (Simple Spectral Access Protocol) service.  

\section{Summary}\label{sec_summary}

We have modified the original HUT data-calibration pipeline to create a new, more user-friendly data product: a time-tagged photon-event list in a format similar to the \fuse\/ Intermediate Data File that is already familiar to many users in the community.  Using this new pipeline, we have reprocessed the entire HUT data archive from both the Astro-1 and Astro-2 Spacelab missions, producing for each exposure an IDF and an extracted spectral file in a modern FITS format  with all relevant supporting information in the FITS file headers.  We have retrieved HUT TV guider images from our archives, converted them to FITS format, and populated their headers with complete astrometric information.  We have produced quick-look plots for each exposure to provide an accessible way for users to pre-screen the newly processed data.  We also provide software with which users may extract a fully-calibrated spectrum from the archived IDFs in the event that settings different from the default are desired.  All of these files and software, together with documents, photographs, and other supporting information, are available from the HUT section of the MAST archive.

\acknowledgments

This work has been  supported by NASA ADP Grant \#NNX09AC70G to the Johns Hopkins University and by the Center for Astrophysical Sciences at the Johns Hopkins University. It has made use of NASA's Astrophysics Data System Bibliographic Services (ADS) and the Mikulski Archive for Space Telescopes (MAST), hosted at the Space Telescope Science Institute.  STScI is operated by the Association of Universities for Research in Astronomy, Inc., under NASA contract NAS5-26555. Support for MAST for non-\hst\/ data is provided by the NASA Office of Space Science via grant NAG5-7584 and by other grants and contracts.
IRAF, the Image Reduction and Analysis Facility, is distributed by the National Optical Astronomy Observatories, which are operated by the Association of Universities for Research in Astronomy, Inc., under cooperative agreement with the National Science Foundation.
IDL is a registered trademark of Exelis Visual Information Solutions, Inc., for its Interactive Data Language software.  

{\it Facilities:} \facility{HUT}.

\appendix

\section{File Formats}\label{sec_formats}

All HUT data are stored as FITS files \citep{Hanisch:01} containing one or more Header + Data Units (HDUs).  The first is called the primary HDU (or HDU~1); it consists of a header and an optional N-dimensional image array.  The primary HDU may be followed by any number of additional HDUs, called ``extensions.''  Each extension has its own header and data unit.  The old HUT pipeline stored spectra in image extensions (a two-dimensional array of pixels); the new pipeline employs binary-table extensions (rows and columns of data in binary representation) for both IDFs and extracted spectra.  The new pipeline uses the \anchor{http://heasarc.gsfc.nasa.gov/docs/software/fitsio/fitsio.html}{CFITSIO} subroutine library \citep{Pence:99} to read and write FITS files.  All of our data products are fully compliant with Virtual Observatory requirements \citep{Tody:2012}.  

\begin{deluxetable}{lll}
\tablecolumns{3}
\tablewidth{0pt}
\tablecaption{Format of Intermediate Data File\label{idf}}
\tablehead{
\colhead{Array Name} & \colhead{Format} & \colhead{Description}
}
\startdata
\cutinhead{Primary Header-Data Unit (HDU 1)}
\multicolumn{3}{l}{Header only.  Keywords contain exposure-specific information.} \\
\cutinhead{HDU 2: Photon Event List}
TIME     &       FLOAT    & Photon arrival time (seconds) \\
XRAW    &        SHORT &   Detector X coordinate (0--2048) \\
X          &     SHORT  &  Corrected for telescope motion \\
COUNTS   &       LONG &   Raw counts per pixel \\
STATUS\_FLAGS    &    BYTE  &   Status flags (see Table \ref{flags}) \\
\cutinhead{HDU 3: Calibration Table}
FLATFIELD   &       DOUBLE  &       Flat-field correction (unitless)  \\
WAVELENGTH    &       DOUBLE  &       Output wavelength array (\AA) \\
INV\_SENS	&	DOUBLE	&	Inverse-sensitivity curve (erg/cm$^2$/\AA) \\
DT\_COR\_FAC	&	DOUBLE	&	Dead-time correction array (unitless) \\
DT\_COR\_ERR	&	DOUBLE	&	Dead-time uncertainties (unitless) \\
\cutinhead{HDU 4: Timeline Table}
TIME     &       FLOAT     &      From raw file headers (seconds) \\
STATUS\_FLAGS &   BYTE     &       Status flags (Table \ref{flags}) \\
LIMBANG   &   FLOAT      &    Limb angle (degrees)  \\
LONGITUDE    &   FLOAT     &     Spacecraft longitude (degrees)  \\
LATITUDE     &   FLOAT       &   Spacecraft latitude (degrees)  \\
HUT\_PITCH\_ERR    &   FLOAT     &      From HUT guide stars (degrees) \\
HUT\_YAW\_ERR    &   FLOAT     &      From HUT guide stars (degrees) \\
SOURCE\_RATE	&	SHORT	&	Excluding airglow lines (counts/second) \\
LY\_ALPHA\_RATE	&	SHORT	&	Lyman $\alpha$ airglow (counts/second) \\
O\_I\_RATE	&	SHORT	&	\oone\ $\lambda 1304$ airglow (counts/second) \\
RATE\_850\_895	&	SHORT	&	Background (850--895 \AA) (counts/second) \\
\cutinhead{HDU 5: IMCS Table}
TIME   &       FLOAT  &       From raw file headers (seconds)  \\
PIXEL\_SHIFT    &       SHORT  &       Correction for pointing error (pixels) \\
IMCS\_PITCH\_ERR	&	FLOAT	&	Computed pitch error (arcsec) \\
IMCS\_YAW\_ERR	&	FLOAT	&	Computed yaw error (arcsec)
\enddata
\tablecomments{Times are relative to the exposure start time, stored in the header keyword EXPSTART.}
\end{deluxetable}

\subsection{Intermediate Data File (IDF)}\label{idf_format}

The Intermediate Data File (IDF; suffix ``idf.fits'') consists of a primary HDU and four FITS binary table extensions; their contents are listed in Table \ref{idf}.  The primary HDU (HDU~1) consists only of a file header; its keywords are gleaned from the raw data files, the HUT engineering database (constructed from data provided by the telescope, the spacecraft, or NASA), and the sequence file for that observation.  Additional keywords are populated by subsequent pipeline routines.  

The first binary-table extension (HDU~2) contains the photon event list.  The TIME and XRAW arrays are taken from the raw data file; photon coordinates corrected for target motion are recorded in the X array.  The COUNTS array lists the number of photon events recorded by each pixel in each raw data file.  Zero-valued pixels are ignored.  The STATUS\_FLAGS, stored as an array of 8-bit bytes (Table \ref{flags}), identify photons that arrived during ``bad'' times (target out of aperture, during an SAA passage, \etc) or are contaminated by airglow features.  Bad photons are not deleted from the IDF, but merely flagged.  For each bit, a value of 0 indicates that the photon is ``good,'' except for the day/night flag, for which 0 = night and 1 = day.  It is possible to modify these flags without re-running the pipeline.  For example, one could exclude day-time photons to reduce the strength of the Lyman $\alpha$ airglow feature.

The second extension (HDU~3) is called the calibration table.  It contains five 2048-element arrays used to calibrate the extracted spectra: the flat-field correction, the wavelength calibration, the inverse-sensitivity curve, the dead-time correction, and its associated error array.  Initially set to zero, these arrays are populated by the various steps of the calibration pipeline.

The third extension (HDU~4) is called the timeline table.  It contains status flags and spacecraft  parameters used by the pipeline.  An entry in the timeline table is created for each raw data file.  The nominal time spacing is thus two seconds, but may be longer if a file is rejected or missing.  The day/night, airglow, limb-angle, SAA, and high-background flags of the STATUS\_FLAGS array are populated when the IDF is created; the out-of-aperture flag is populated later in the pipeline.  The elements of the LIMBANG, LONGITUDE, and LATITUDE arrays are extracted from the HUT engineering database.  The HUT\_PITCH\_ERROR and HUT\_YAW\_ERROR arrays are taken from the headers of the raw data files.  The SOURCE\_RATE, LY\_ALPHA\_RATE, and O\_I\_RATE arrays are computed from spectral information in the raw data files.

The fourth extension (HDU~5), called the IMCS table, contains pointing information computed by the program {\tt imcs\_ph\_corr}, described in \S\ \ref{sec_imcs}.  This program uses its own criteria to select ``good'' raw data files, so its TIME array may not match those of the photon event list and timeline table.  The PIXEL\_SHIFT array contains the (integral) shift in pixels required to correct for target offsets from the center of the aperture.  This correction is additive.  The IMCS\_PITCH\_ERR and IMCS\_YAW\_ERR arrays contain the offset of the target from the aperture center as computed by {\tt imcs\_ph\_corr}.  If the program fails to compute a solution (for example, if sufficient pointing information is unavailable), then this extension will not be present.

Extracted spectra files files (\S\ \ref{spec_format}) employ the standard FITS binary table format, listing WAVE, FLUX, ERROR, \etc\ for each pixel in turn. The intermediate data files have a slightly different format, listing all of the photon arrival times, then the raw X coordinates, then the corrected X coordinates. Formally, the table has only one row, and each element of the table is an array.  (To use the STSDAS terminology, IDFs are written as 3-D tables.)  The MDRFITS function from the \anchor{http://idlastro.gsfc.nasa.gov/}{IDL Astronomy User's Library} \citep{Landsman:93} can read both file formats; some older FITS readers cannot. 

\begin{deluxetable}{lll}
\tablecolumns{3}
\tablewidth{0pt}
\tablecaption{Format of Extracted Spectral Files\label{spec}}
\tablehead{
\colhead{Array Name} & \colhead{Format} & \colhead{Description}
}
\startdata
\cutinhead{Primary Header-Data Unit (HDU 1)}
\multicolumn{3}{l}{Header only.  Keywords contain exposure-specific information.} \\
\cutinhead{HDU 2: Extracted Spectrum}
WAVE     &       FLOAT    & Wavelength (\AA) \\
FLUX     &       FLOAT    & Flux (\flux) \\
ERROR     &       FLOAT    & Poisson error (\flux) \\
COUNTS     &       INT    & Raw counts, corrected for motion \\
WEIGHTS     &       FLOAT    & Fully-corrected counts \\
\cutinhead{HDU 3: Orbital-Night Spectrum}
WAVE     &       FLOAT    & Wavelength (\AA) \\
FLUX     &       FLOAT    & Flux (\flux) \\
ERROR     &       FLOAT    & Poisson error (\flux) \\
COUNTS     &       INT    & Raw counts, corrected for motion \\
WEIGHTS     &       FLOAT    & Fully-corrected counts
\enddata
\end{deluxetable}

\subsection{Extracted Spectral Files}\label{spec_format}

Extracted spectra (suffix ``cal.fits''; \S\ \ref{sec_extract}) are stored in a pair of binary-table extensions.  The first extension contains a spectrum derived from the entire exposure.  The second extension contains a spectrum derived only from the orbital-night data.  Two extensions are always present; if no night data are available, the second extension contains a null-valued spectrum.  If the entire exposure was obtained during orbital night, the two spectra are identical.  Exposure times and photometric corrections for both spectra are stored in the primary header.
 
The spectral-file format is presented in Table \ref{spec}.  Spectra maintain the 2048 pixels of the raw data.  The WAVE array records the  wavelength of each pixel.  The COUNTS array contains the raw counts in each pixel, corrected only for telescope motions.  The WEIGHTS array is fully corrected for detector effects, scattered light, and second-order contamination.  The FLUX array is the product of the WEIGHTS and the inverse-sensitivity curve.  The ERROR array is computed assuming Poisson statistics; no estimate of systematic errors is included.

\subsection{HUT Video Files}\label{video_format}

Images of the target field taken with the HUT TV camera (\S\ \ref{sec_video}) are stored as FITS images with the suffix ``tv.fits.''  Each file consists of a single HDU.  File header keywords are copied from the associated IDF, and a complete set of astrometric keywords is appended.  

{\em Caveats:}  A TV camera image could be transmitted as a single image or as eight sub-frames, each downloaded as a separate file.  In the latter case, the final image is a composite of sub-frames taken as much as a minute apart.  If the telescope was moving during that time, then a single star may appear in two or more sub-frames.  The file-header keywords reflect the episodic nature of the data: the exposure start and stop times represent the times of the initial and final sub-frame, respectively, and the quoted exposure time represents the integration time for each sub-frame.

\section{Notes on Individual Targets}\label{sec_notes}

{\em Comet Levy:\/}  Because Comet Levy (1990c) was observed entirely during orbital day, its spectrum is dominated by emission from the earth's atmosphere \citep{Feldman:91}.  To allow a determination of the airglow contribution, the comet was observed for 10 minutes, then the IPS was offset by approximately 2\arcmin.  After five minutes, the IPS was returned to its initial position.  Following \citeauthor{Feldman:91}, we have divided the observation into three exposures (HUT118001-3) with target names C-LEVY, C-LEVY\_OFFSET, and C-LEVY and exposure times of approximately 600, 300, and 300 s, respectively.  In our quick-look plots, we replace the standard source-rate plots with plots of the Lyman $\alpha$ count rate.

{\em Jupiter:\/}  The Jupiter system was observed on both Astro missions.  Six exposures were obtained on Astro-1 \citep{Feldman:93}:  On 1990 December 7, Jupiter was acquired through the 17\arcsec\ $\times$ 116\arcsec\ aperture (HUT106901, JUPITER); the slit was then centered on the west ansa of the Io torus and aligned parallel to the plane of the torus (HUT106902, IO\_TORUS); the 9\farcs4 $\times$ 116\arcsec\ aperture was moved into place and the planet centered in the aperture (HUT106903, JUPITER); finally, to obtain an auroral spectrum, the telescope was offset to a position near the south polar limb of the planet (HUT106904, JUPITER\_AURORA).  On 1990 December 10, the 9\farcs4 $\times$ 116\arcsec\ aperture was used to observe the east ansa of the Io taurus (HUT103705, IO\_TORUS), then centered on the planet (HUT103706, JUPITER).

Astro-2 produced 11 exposures of Jupiter and its environs: 
two exposures centered on the Jupiter equator (HUT204401-2, JUPITER\_EQUATOR);  
one centered on the leading hemisphere of Io (HUT205503-4, IO\_EAST);
one with the 10\arcsec\ $\times$ 56\arcsec\ aperture aligned with the Io torus (HUT202405, IO\_TORUS);
one centered on the north polar region (HUT218706, JUPITER\_AURORA);
three mapping the Io torus with the 20\arcsec\ aperture  (HUT216507-9, IO\_TORUS\_POS1-3); and
two centered on the trailing hemisphere of Io (HUT217910-11, IO\_WEST).

{\em NGC 1535:\/}  The Astro-1 observation of the planetary nebula NGC 1535 \citep{Bowers:95} was intended to begin with the central star centered in the aperture (HUT112301, NGC1535\_CSPN), then move by 9\arcsec\ to obtain a spectrum of the planetary nebula (HUT112302, NGC1535\_NEB).  Pointing jitter caused the nebular spectrum to be contaminated by the central star at intervals throughout the second exposure.  For this exposure, we modified our target-out-of-aperture routine to reject times of elevated count rate.

{\em NGC 7023:\/}  One star within this open cluster, HD 200775, was observed during Astro-1.  On Astro-2, a reflection nebula, offset from the star by -50\farcs 2 in right ascension and and -81\farcs 6 in declination, was observed several times.  The nebular exposures are labeled NGC7023\_OFFSET.

{\em $\rho$ Ophiuchi A+B (HD~147933):\/}  The original intent was to observe both components of this binary system, then to offset by +45\arcsec\ in right ascension to the reflection nebula IC 4604.  The stellar observation was sacrificed to allow for more night time on the nebula.  The resulting exposures are labeled RHOOPHAB\_OFFSET.

{\em $\xi$ Persei (HD~24912):\/}  With an apparent magnitude of 4.04, this blue giant was too bright for HUT to observe directly.  Instead, the HUT mirror was tilted to observe an \htwo\ region 100\arcsec\ to the north.  Both of the resulting exposures are thus labeled XI-PER\_OFFSET.

{\em White Dwarfs:\/}  As discussed in \S\ \ref{sec_fluxcal}, no inverse-sensitivity curve is available for the second-order spectra of G191-B2B (HUT109702) and HZ~43 (HUT105202) obtained on Astro-1.  Both their extracted spectra and the quick-look plots derived from them have units of counts per second.




\end{document}